\documentclass[12pt]{article}
\usepackage[utf8x]{inputenc}
\usepackage[T1]{fontenc}
\usepackage{multicol}
\usepackage{float}
\usepackage{amssymb}
\usepackage{amsthm}
\usepackage{bbold}
\usepackage{latexsym}
\usepackage{cite}
\usepackage{amsmath}
\usepackage{amsfonts} \parindent 0pt \parskip.2cm \topmargin -1.0cm
\oddsidemargin=0.25cm\evensidemargin=0.25cm \newfont{\bbbold}{msbm10
  scaled \magstep1}

 \newcommand{\fp}{\frac1{(2\pi)^{\frac32}}}

 \newcommand{\m}{{\mu}}

\def\pslash#1{{\setbox0=\hbox{$#1$}
    \rlap{\ifdim\wd0>.7em\kern.22\wd0\else\kern.1\wd0\fi /}#1}}

\def\be{\begin{equation}} \def\ee{\end{equation}}
\def\ba{\begin{array}} \def\ea{\end{array}}

\begin{document}
\begin{flushright}
\end{flushright}
\thispagestyle{empty}
\begin{center}
{\Large\sf Preconjugate variables in quantum field theory and their use}\\[10pt]
\vspace{.5 cm}

\vspace{1.0ex}

Albert Much$^{\star}$ \footnote{e-mail: a.much`at'gmx.at},
Steffen Pottel$^{\natural\dagger}$\footnote{e-mail: pottel`at'mis.mpg.de},
Klaus Sibold$^{\dagger}$\footnote{e-mail: sibold`at'physik.uni-leipzig.de}\\ 
$^{\star}${Instituto de Ciencias Nucleares,\\ UNAM, M\'exico D.F. 04510,\\
M\'exico} \\

$^{\natural}$Max-Planck Institute for Mathematics in the Sciences\\
Inselstr.\ 22\\
D-04103 Leipzig\\
Germany\\

$^{\dagger}$Institut f\"ur Theoretische Physik\\
Universit\"at Leipzig\\
Postfach 100920\\
D-04009 Leipzig, Germany

\end{center}

\vspace{2.0ex}
\begin{center}
{\bf Abstract}
\end{center}

\noindent
Preconjugate variables $X$ have commutation relations with the energy-momentum
$P$ of the respective system which are of a more general form than just the
Hamiltonian one. Since they have been proven useful in their own right for 
finding new spacetimes we present here a study of them. Interesting examples
can be found via geometry: motions on the mass-shell for massive and massless
systems, and via group theory: invariance under special conformal transformations
of mass-shell, resp.\ light-cone -- both find representations on Fock space.
We work mainly in ordinary fourdimensional
Minkowski space and spin zero. The limit process from non-zero to vanishing
mass turns out to be non-trivial and leads naturally to wedge variables.
We point out some applications and extension to more general spacetimes.
In a companion paper we discuss the transition to conjugate pairs.\\

\newpage
\section{Introduction}

\noindent
It is current belief that the unification of gravity with the successful
models of particle physics requires new structures: avoiding a clash of
measurement with the microscopic structure of spacetime at distances
of the Planck scale might be possible by introducing coordinate operators $Q_\mu$
as non-commuting observables \cite{DoplicherFredenhagenRoberts}. One way of
obtaining such objects is via the construction of conjugate pairs $\{P_\mu,Q_\nu\}$
\be\label{cp}
[P_\mu,Q_\nu]= i\eta_{\mu\nu},
\ee
where $P$ is to be identified with the total energy-momentum operator of
the system in question.\\
In \cite{Sibold, SiboldSolard,EdenSibold} respective studies have been
started. It turned out that in practice one better finds first
{\sl preconjugate} pairs satisfying
\be\label{precp}
[P_\mu,X_\nu] = iN_{\mu\nu},
\ee
with $N_{\mu\nu}$ denoting an operator which can be inverted in a sense
to be specified. Depending on respective state spaces this may be 
mathematically delicate and rendering ``operators'' to bilinear forms or
the like.\\
We do not attempt to characterize possible $Q$'s axiomatically, however
rely on geometrical and group theoretical notions for finding suitable
candidates. Suppose we choose Fock space as state space. Then one can
interpret creation and annihilation operators as mappings of the manifold
$p^2=m^2$ -- the mass shell, to Hilbert space. A motion on the manifold, 
i.e. a motion which does not leave the manifold, can then be described
in Hilbert space by a charge like operator: this will be a candidate
for $X$. If $m^2=0$, the lightcone $p^2=0$ is left invariant by
infinitesimal special conformal transformations, the respective charge is
also a candidate for an $X$. Similarly the light cone $x^2=0$ in, say four
dimensional Minkowski space, is also left invariant by infinitesimal
special conformal transformations, also represented in Hilbert space by 
charge like operators. Here one quickly realizes that group theoretic 
considerations, apriori not confined to a state space specified in
advance, will be a useful tool. In a nutshell, these are the main
topics to be treated in the present paper. We will analyze (\ref{precp})
and leave the study of inversion to achieve (\ref{cp}) 
to a subsequent paper \cite{PottelSiboldII}.\\
In sect.2 we work out the details of this program on-shell, in
sect.3 we employ Green functions, i.e.\ work off-shell. In sect.4 we
consider a generalization to other spacetimes (essentially (Anti-)deSitter))
and discuss an application of a preconjugate pair via deformation theory.
In sect.5 we summarize our results and mention open questions.\\

\newpage
\section{On-shell approach}
\subsection{One-particle wave functions}
To begin with we realize preconjugate operators $X_\mu$ as differential operators acting on one-particle
wave functions. For simplicity we treat here only the scalar case. Below we shall discuss 
generalizations.\\
Starting point is the Klein-Gordon equation with either $m^2 \not=0$ or $m^2=0$. I.e.\
we consider square integrable functions $f$ from $\mathbb{R}^4 \rightarrow \mathbb{C}$ and
impose the equation
\begin{equation}\label{KleinGordon}
	(\Box + m^2)f(x) = 0 \qquad m^2 \not= 0 \quad {\hbox{\rm or}}\quad  m^2=0
\end{equation}
as additional constraint. We realize the translations as before by
\begin{equation}
	P_\mu f(x) = -i\frac{\partial}{\partial x^\mu}f(x)  \qquad  P_\mu\tilde{f}(p)
	= p_\mu \tilde{f}(p)
\end{equation}
and discuss now preconjugate partners to them as they are suggested by geometry of
the manifold $p^2 = m^2$ or the invariance of (\ref{KleinGordon}).\\

\subsubsection{ The differential operator $X(\nabla)$}

\noindent
Let us define (massive case) a differential operator $X_\nu$ by
\begin{eqnarray}\label{Xnabladiff}
	X_\nu &=& i\nabla_\nu \equiv i(\frac{\partial}{\partial p^\nu}
	-\frac{p_\nu}{m^2}p^\lambda\frac{\partial}{\partial p^\lambda})\qquad\mbox{in mom.\ space},\\
	X_\nu &=& -( x_\nu + \frac{5}{m^2}\frac{\partial}{\partial x^\nu} + 
	\frac{1}{m^2}x^\lambda\frac{\partial^2}{\partial x^\lambda\partial x^\nu})\qquad\mbox{in pos.\ space}.
\end{eqnarray}
Since it satisfies
\be\label{PXnabladiff}
[P_\mu,X_\nu] = i(\eta_{\mu\nu} - \frac{p_\mu p_\nu}{m^2}),
\ee
it clearly qualifies as an operator preconjugate to $P_\mu$, the latter being interpreted as
momentum operator. Obviously Poincar\'e covariance is manifest. In fact, $\nabla_\nu$ is
known as a {\sl tangential derivative} \cite{ColemanMandula} and maintains the mass shell condition
resulting from the Klein-Gordon equation.\\
For the commutator of $X$'s one finds
\begin{eqnarray}\label{comXnabladiff}
[X_\mu,X_\nu]\tilde{f} &=& \frac{-1}{m^2}(p_\mu\frac{\partial}{\partial p^\nu} 
	                               - p_\nu\frac{\partial}{\partial p^\mu})\tilde{f}\\
				       &=& \frac{i}{m^2}M_{\mu\nu}\tilde{f},
\end{eqnarray}
i.e.\ the Lorentz transformations: the preconjugate $X(\nabla)$ is non-commutative.\\
One technical remark is in order. If in solutions $\tilde{f}$ of the Klein-Gordon
equation the component
$p_0$ of the argument is replaced by its values $p_0=\pm \omega_p \equiv \pm
\sqrt{m^2 - p^lp_l}$ then
the derivative $\partial/\partial p_0$ is identically zero and the sum of
derivatives
$p^\lambda \partial/\partial p^\lambda$ is reduced accordingly to
$p^l \partial/\partial p^l$.
I.e.\ the motions induced by $X_\nu$ are confined to a 3-dimensional
submanifold in $p$-space, hence also in $x$-space: Energy, respectively time are
fixed and the preconjugate pairs reduce to
those of purely space-type.
This is best seen in the rest system, ${\bf p}=0$, clearly $X_0=0$.\\
These algebraic properties originate from geometry, as we now show. Embedding the hyperboloid $p_0=\omega_p$
into flat Minkowski space ${\mathbb{R}}^4$ with metric $\eta = diag(+1,-1,-1,-1)$ we achieve to
end up at a submanifold with induced metric $g$ as follows:
\begin{eqnarray}\label{inducedmetric} 
\mbox{choose variables} \qquad q_j &=& p_j \qquad j=1,2,3\\
\mbox{identify submetric} \qquad g({\bf q})_{jk} &=& \frac{\partial p^\mu}{\partial q^j}
	                                            \frac{\partial p^\nu}{\partial q^k}\eta_{\mu\nu}\\
\mbox{find submetric} \qquad    g({\bf q})_{jk} &=& \eta_{jk} + \frac{q_jq_k}{m^2 + q_lq_l} \qquad j,k,l=1,2,3\\
\mbox{find inverse submetric} \qquad g({\bf q})^{jk} &=& \eta^{jk} - \frac{q^jq^k}{m^2} 
\end{eqnarray}

We note first that there holds
\be\label{subnabla}
\nabla^j=(\eta^{jk}-\frac{p^jp^k}{m^2})\frac{\partial}{\partial p^k}
                                    =g^{jk}\partial_k
\ee
(with obvious relation to (\ref{Xnabladiff})). We then calculate the Christoffel
symbols for the submetric
\be
\Gamma^j_{kl}=\frac{p^j}{m^2}g_{kl}
\ee
and finally verify that
\be\label{subnablcons}
	\nabla_r g^{jk}=0.
\ee
Hence the Christoffel symbols constitute the Levi-Civita connection,
the $\nabla$'s are covariant derivatives and all quantities are intrinsically
defined, i.e.\ not a property of the chosen parametrization, but of the
(sub)manifold involved.\\


\subsubsection{The differential operator $X(<)$} 
This operator originates from $X(\nabla)$ by going over to ``wedge coordinates''
(``$<$'' for ``wedge''). First, it is an operator in its own right, but
second, it
will eventually suggest the limit of vanishing mass in a controllable way.
It is obvious that for $X(\nabla)$ this limit does not exist.\\
In order to proceed we first introduce new variables in $p$-space by
\begin{eqnarray}\label{wedgevariablesp}
	p_u&=&\frac{1}{\sqrt{2}}(p_0-p_1) \qquad\qquad p_0=\frac{1}{\sqrt{2}}(p_v+p_u)\\
	p_v&=&\frac{1}{\sqrt{2}}(p_0+p_1) \qquad\qquad p_1=\frac{1}{\sqrt{2}}(p_v-p_u).
\end{eqnarray}
For later use we also note the respective variable change in $x$-space:
\begin{eqnarray}\label{wedgevariablesx}
	u&=&\frac{1}{\sqrt{2}}(x^0-x^1) \qquad\qquad x^0=\frac{1}{\sqrt{2}}(v+u)\\
	v&=&\frac{1}{\sqrt{2}}(x^0+x^1) \qquad\qquad x^1=\frac{1}{\sqrt{2}}(v-u).
\end{eqnarray}
Note: $p^u=p_v, p^v=p_u$.
The mass shell condition is given by
\be\label{wedgemassshell}
2p_up_v-p_ap_a = m^2 \qquad a=2,3 \qquad\qquad{\mbox{summation over a}}
\ee
We call these variables ``wedge'' variables; in the case of $m^2=0$ ``light wedge''
variables.\\
The geometry is governed by metric tensors which we now study. In a first step we
go over from variables $p_\mu$ to variables $p'_\mu$ given by
\be
p_0'=p_u \qquad p_1'=p_v \qquad p_a'=p_a \qquad a=2,3.
\ee
The new metric $g^{\mu\nu}(p')$ is obtained from the old metric $\eta^{\mu\nu}=diag(+1,-1,-1,-1)$
by
\be\label{wedgemetricdef}
g^{\mu\nu}(p')=\frac{\partial p_\lambda}{\partial p'_\mu}\frac{\partial p_\rho}{\partial p'_\nu}\eta^{\lambda\rho}
\ee
and given by
\be\label{wedgemetric}
g^{\mu\nu}(p')=\left(\begin{array}{cccc}
		          0&1&0&0\\  
                          1&0&0&0\\
                          0&0&-1&0\\
                          0&0&0&-1\\ \end{array}\right) \equiv\bar{\eta}^{\mu\nu}.
\ee
The inverse metric $g_{\mu\nu}(p')=(g^{-1})^{\mu\nu}(p')$ has the same form:
\be\label{inversewmetric}
g_{\mu\nu}(p')=\left(\begin{array}{cccc}
		          0&1&0&0\\  
                          1&0&0&0\\
                          0&0&-1&0\\
                          0&0&0&-1\\ \end{array}\right) \equiv\bar{\eta}_{\mu\nu}.
\ee
The basis transformation $p\rightarrow p'$ was linear, hence flat space goes into flat
space, the metric $\eta_{\mu\nu}$ into $\bar{\eta}_{\mu\nu}$.\\

\noindent
In the second step we implement the (non-linear) mass shell constraint $p'_0=(m^2+p'_ap'_a)/2p'_1$,
introduce three variables $q_j$ and define the metric $g(q)^{ij}$ of the submanifold
by
\be\label{subuvconstraint}
\begin{array}{ccccc}
	q_1&=&p'_1&=&p_v\\
	q_2&=&p'_2&=&p_2\\
        q_3&=&p'_3&=&p_3\\ \end{array} \qquad g(q)^{ij}=
	      \frac{\partial p'_\mu}{\partial q_i}\frac{\partial p'_\nu}{\partial q_j}\bar{\eta}^{\mu\nu}.
\ee
The result for the metric reads
\begin{align}\label{subuvmetric}
g(q)^{ij} = \left(\begin{array}{ccc}
		        -\frac{m^2+q_aq_a}{q_1^2}&\frac{q_2}{q_1}&\frac{q_3}{q_1}\\
		        \frac{q_2}{q_1}&-1&0\\
\frac{q_3}{q_1}&0&-1\\ \end{array}\right)^{ij},\qquad
&g(p_v,p_a)^{ij} = \left(\begin{array}{ccc}
		        -\frac{m^2+p_ap_a}{p_v^2}&\frac{p_2}{p_v}&
		                                 \frac{p_3}{p_v}\\
		        \frac{p_2}{p_v}&-1&0\\
				 \frac{p_3}{p_v}&0&-1\\ \end{array}\right)^{ij}.
\end{align}
For the inverse metric $g(q)_{ij}$ we obtain
\begin{align}\label{inverseuvmetric}
g(q)_{ij} = -\frac{q_1^2}{m^2}\left(\begin{array}{ccc}
                                 1&\frac{q_2}{q_1}&\frac{q_3}{q_1}\\
	           \frac{q_2}{q_1}&\frac{m^2+q_2q_2}{q_1^2}&\frac{q_2q_3}{q_1^2}\\
\frac{q_3}{q_1}&\frac{q_2q_2}{q_1^2}&\frac{m^2+q_3q_3}{q_1^2}\\ \end{array} \right)_{ij},\qquad
&g(p_v,p_a)_{ij} = -\left(\begin{array}{ccc}
		\frac{p_v^2}{m^2} &\frac{p_vp_2}{m^2}&\frac{p_vp_3}{m^2}\\
	           \frac{p_2p_v}{m^2}&\frac{m^2+p_2p_2}{m^2}&\frac{p_2p_3}{m^2}\\
\frac{p_3p_v}{m^2}&\frac{p_2p_2}{m^2}&\frac{m^2+p_3p_3}{m^2}\\ \end{array} \right)_{ij}.
\end{align}

\noindent
Our next task is to find the tangential derivatives $\nabla(<)$ to the mass shell $2p_up_v - p_ap_a = m^2$.
They are defined by the requirement
\be\label{tangderivuv}
\nabla (2p_up_v-p_ap_a)= 0 \qquad \mbox{at}\quad  p_u=\frac{m^2+p_ap_a}{2p_v}
\ee
Returning to the wedge variables $p$ we define in analogy to the Coleman/Mandula
$\nabla$
\begin{eqnarray}\label{tangderans}
\nabla^u = \frac{\partial}{\partial p_u} - \frac{p^u}{m^2}p_\lambda\frac{\partial}{\partial p_\lambda}
&\qquad&
\nabla^2 = \frac{\partial}{\partial p_2} - \frac{p^2}{m^2}p_\lambda\frac{\partial}{\partial p_\lambda}\\	
\nabla^v = \frac{\partial}{\partial p_v} - \frac{p^v}{m^2}p_\lambda\frac{\partial}{\partial p_\lambda}
&\qquad&
\nabla^3 = \frac{\partial}{\partial p_3} - \frac{p^3}{m^2}p_\lambda\frac{\partial}{\partial p_\lambda}.
\end{eqnarray}

Although in the wedge variables both shells of the hyperboloid are covered
the transition from standard to wedge variables is essentialy a change of
parametrization only and again the respective $\nabla$'s will express
generic properties of the underlying submanifold.\\
At the level of one-particle wave functions the respective operators $X(<)$, 
defined by 
\be\label{nablauvdef}
X(<)= i\nabla(<) \qquad \hbox{\rm with indices}\,\,  \{u,v,a=1,2\}
\ee 
do not seem to be significantly different from the previous $X(\nabla)$. 
It is, however, to be expected that on $n$-particle
wave functions (or on $n$-particle Fock space states) they permit an
interesting distinction between centre of mass and relative
variables (s.\cite{Heinzl}).

\subsubsection{The differential operator X(light wedge)$\equiv$ X($<_0$)}
Here we study the massless limit of $X(<)$ denoted by $X(<_0)$.\\
We observe that for the submetric (\ref{subuvmetric}) the limit
$m^2\rightarrow 0$ exists, but
that ``coordinate'' singularities appeared; for the inverse submetric
(\ref{inverseuvmetric}) the limit of
vanishing mass does not exist, however no ``coordinate'' singularities
occur. (The determinant of (\ref{subuvmetric}) vanishes for $m^2=0$.)\\
If we wish to avoid both of these shortcomings we have to perform a
substantial change in our setting. We propose the following one: we
consider the subspaces $\{p_u,p_v\}$ and $\{p_2,p_3\}$ as being 
independent from each other. Then in (\ref{subuvconstraint}) the range of
$\{\mu,\nu\}$
is restricted in the summation accordingly to $\{u,v\}$, resp.\ $\{2,3\}$,
hence the off-diagonal terms in (\ref{subuvmetric},\ref{inverseuvmetric})
disappear. The only remaining relation between the two sectors is
the mass-shell constraint
\be\label{mass0constr}
2p_up_v = p_2p_2+p_3p_3 \equiv p_ap_a.
\ee
For the induced submetric we end up with
\be\label{1,1;0,2metric}
g(p_v,p_a)^{ij}=\left(\begin{array}{ccc}
	          -\frac{p_ap_a}{p_v^2}&0&0\\
	   0&-1&0\\
	0&0&-1
\end{array}\right),
\ee
for the induced inverse submetric with
\be\label{1,1;0,2invmetric}
g(p_v,p_a)_{ij}=\left(\begin{array}{ccc}
                 -\frac{p_v^2}{p_ap_a}&0&0\\
	   0&-1&0\\
	0&0&-1
\end{array}\right).
\ee
Neither of the two expressions $p_v^2$, nor $p_ap_a$ must vanish.
This is no serious restriction because it only means that the 
circle $p_ap_a={\rm const}$ must have non-vanishing radius and 
the hyperbola $2p_up_v=p_ap_a={\rm const}\not=0$ has its standard
singularity in $p_v=0$ which has to be taken care off.\\
As far as interpretation is concerned it is clear that we went from a
four dimensional momentum space carrying the manifold $p^2=m^2$
and associated Lorentz transformations $SO(1,3)$ to a direct product space
with two two-dimensional factors carrying the manifolds
$2p_up_v={\rm const}$, resp. $p_ap_a={\rm const}$ and thus the invariance
groups $SO(1,1)$ resp.\ $SO(2)$.\\
It is to be noted that here we can cover both shells of the hypercone
$p^\mu p_\mu=m^2=0$ if we permit $p_0=\pm\omega_p$.\\
The differential operators to be found for vanishing mass should be
tangential to the respective submanifolds. Hence they are defined by
the requirement
\be\label{lightwedgetangder}
\nabla (2p_up_v-p_ap_a)= 0 \qquad \mbox{at}\quad  p_u=\frac{p_ap_a}{2p_v}
\ee
Again, in terms of wedge variables $p$ we try in analogy to the
Coleman/Mandula $\nabla$ the ansatz
\begin{eqnarray}\label{0tangderans}
\nabla^u = \frac{\partial}{\partial p_u} - \frac{p_v}{2p_up_v}(p_v\frac{\partial}{\partial p_v}
                                        	+p_u\frac{\partial}{\partial p_u})&\quad&      
  \nabla^2 = \frac{\partial}{\partial p_2} +\frac{p^2}{\tilde{\omega}_p^2}p_b\frac{\partial}{\partial p_b}\\    
\nabla^v = \frac{\partial}{\partial p_v} - \frac{p_u}{2p_up_v}(p_v\frac{\partial}{\partial p_v}
                                                    +p_u\frac{\partial}{\partial p_u})&\quad&      
   \nabla^3 = \frac{\partial}{\partial p_3} + \frac{p^3}{\tilde{\omega}_p^2}p_b\frac{\partial}{\partial p_b}.
\end{eqnarray}
Differing from (\ref{tangderans}) the denominators $2p_up_v$ and
$(\tilde{\omega}_p)^2=(\sqrt{p_ap_a}\,)^2$ (sum over $a=2,3$) represent
the curvatures in the respective submanifolds. \\
Using $p^u=p_v$ (\ref{wedgevariablesp}) the expressions for $\nabla^u, \nabla^v$ simplify to
\be\label{0tangderuv}
\nabla^u=\frac{1}{2}(\frac{\partial}{\partial p_u}-\frac{1}{p_u}p_v\frac{\partial}{\partial p_v})\qquad
\nabla^v=\frac{1}{2}(\frac{\partial}{\partial p_v}-\frac{1}{p_v}p_u\frac{\partial}{\partial p_u})
\ee

Those of $\nabla^2, \nabla^3$ can be rewritten as
\be\label{0tangder23}
\nabla^2 = \frac{\partial}{\partial p_2} -\frac{p^2}{p_ap^a}p_b\frac{\partial}{\partial p_b} \qquad
\nabla^3 = \frac{\partial}{\partial p_3} -\frac{p^3}{p_ap^a}p_b\frac{\partial}{\partial p_b}.
\ee
All $\nabla$'s are indeed tangential. \\
Altogether these differential operators satisfy the algebra
\begin{align}\label{nablauvalg}
[\nabla^u,\nabla^v]&=\frac{-1}{2p_up_v}(p_u\frac{\partial}{\partial p^v} 
                             -p_v\frac{\partial}{\partial p^u})=
             \frac{1}{p_ap^a}(p_u\frac{\partial}{\partial p^v} 
                                      -p_v\frac{\partial}{\partial p^u})\\
	[\nabla^u,\nabla^2]&=[\nabla^u,\nabla^3]=[\nabla^v,\nabla^2]=[\nabla^v,\nabla^3]=0\\
[\nabla^2,\nabla^3]&=-\frac{1}{\tilde{\omega}_p^2}
 (p^2\frac{\partial}{\partial p_3}-p^3\frac{\partial}{\partial p_2})=
            \frac{1}{2p_up_v}(p^2\frac{\partial}{\partial p_3}
                                      -p^3\frac{\partial}{\partial p_2}).
\end{align}
Let us note:\\
(1) Due to the respective metrics in the subspaces an $SO(1,1)$ is realized in the $(p_u,p_v)$-variables, whereas
in the $(p_a, a=2,3)$-variables an $SO(2)$ shows up. Again, the factors $1/(2p_up_v)$ and
$1/\tilde{\omega}_p^2$ resp.\ correspond to curvature values in the given subspaces.
As a consequence of the mass shell constraint their values can be considered as constant when
applying the differentiations.\\
(2) The groups $SO(1,1),\, SO(2)$ may be considered as subgroups of the
ambient $SO(1,3)$ which operates in the
ambient four-dimensional Minkowski space.\\

\noindent
In order to familiarize ourselves with these light-wedge-$\nabla$'s
we calculate their action on the components of $p$
\be\label{wedgeuvpupvdiff}
\begin{array}{ccccccccccc}
	        \nabla^v p_u &=&\nabla_u p_u &=&-\frac{p_u}{2p_v} &\qquad& \nabla^v p_v&=&\nabla_u p_v&=&\frac{1}{2}\\
	        \nabla^u p_u &=&\nabla_v p_u &=&\frac{1}{2} &\qquad& \nabla^u p_v&=&\nabla_v p_v&=&-\frac{p_v}{2p_u}.
\end{array}
\ee
Analogously one has in the sector $(a,b)\quad a,b=2,3$
\be\label{wedgexapbdiff}
\begin{array}{ccccccc}
\nabla^2 p_2 &=&1-\frac{p^2p_2}{p_ap^a} &\quad&
                                	   \nabla^2 p_3&=&-\frac{p^2p_3}{p_bp^b}\\
\nabla^3 p_2 &=&-\frac{p^3p_2}{p_bp^b}  &\quad&
	                                   \nabla_3 p_3&=& 1-\frac{p^3p_3}{p_bp^b}.
\end{array}
\ee
We also note the relations
\be\label{lwedgeproject}
p_u\nabla^u+p_v\nabla^v =0 \qquad\qquad p_2\nabla^2+p_3\nabla^3=0.
\ee
They express projection properties, as will become clear below.\\
We may now define operators $X(<_0)$ acting on functions $\tilde{f}(p_u,p_v,p_2p_3)$ as
differential operators by
\begin{eqnarray}
	        X^u(<_0)&=&i\nabla^u  \qquad X^2(<_0)=i\nabla^2\\
	X^v(<_0)&=&i\nabla^v  \qquad X^3(<_0)=i\nabla^3.
\end{eqnarray}
Their algebra is given by
\begin{align}\label{0wedgealg}
	        [X^u(<_0),X^v(<_0)]&=i\frac{1}{P_aP^a}M^{uv}\\
	        [X^2(<_0),X^3(<_0)]&=i\frac{1}{2P_uP_v}M^{23}\\
	        [X^u(<_0),X^2(<_0)]&=[X^u(<_0),X^3(<_0)]=[X^v(<_0),X^2(<_0)]=[X^v(<_0),X^3(<_0)]=0.
\end{align}
Here we have incorporated the mass shell constraint by rewriting
appropriately the denominators.\\
Their commutation relations with the energy-momentum operator $P$ generalize 
(\ref{wedgeuvpupvdiff},\ref{wedgexapbdiff}):
\begin{align}\label{0wedgecpcom}
	[P_\alpha,X_\beta(<_0)]&=\frac{-i}{2}\left(\begin{array}{cc}
                         \frac{-P_u}{P_v} &1\\
                   1 &\frac{-P_v}{P_u}
		  \end{array}\right)    \quad \alpha,\beta=u,v \\
		  [P_a,X_b(<_0)]&=-i\left(\begin{array}{cc}
                              1+\frac{P_2P_2}{P_bP^b}&\frac{-P_2P_3}{P_bP^b}\\
                    \frac{-P_3P_2}{P_bP^b} &1+\frac{P_3P_3}{P_bP^b}
			\end{array}\right)  \quad a,b=2,3
\end{align}
It fits to the before mentioned projection properties (\ref{lwedgeproject}) that the subdeterminants
in the $(u,v)$, resp.\ $(a,b)$ sectors both vanish.\\

\subsubsection{The differential operator $X(\omega)$}

In paper \cite{SiboldSolard} a preconjugate
operator $X^{SiSo}$  
has been proposed whose spatial components are equivalent to the
Newton/Wigner operator
on one-particle states, hence cannot be completed to a four vector. Ad hoc
it had been provided with a ``zero component''. In order to clarify more
completely the surrounding of that operator $X^{SiSo}$ we amend now the zero
component given there by spatial one's. As differential operator on
one-particle wave functions it reads
\be\label{Xomegadif}
X_\nu^{(\omega)} = i(-\frac{p_\nu}{{\omega_p}^2}p^l\frac{\partial}{\partial p^l}) \qquad \omega_p^2 \equiv -p^lp_l
\ee
The zero component of $X(\omega)$ transforms thus as a vector
w.r.t.\ Lorentz transformations, however the spatial components do not
properly transform under boosts.\\
The algebra of these operators is given by
\be\label{omegaalg}
[P_\mu,X_\nu^{(\omega)}]=i\frac{p_\mu p_\nu}{\omega_p^2}\qquad [X_\mu^{(\omega)},X_\nu^{(\omega)}]=0
\ee
All these relations are on-shell, hence $p_0=\omega_p$ and $\partial/\partial p_0 \equiv 0$.
The operators $X^{(\omega)}$ cause some motion in $p$-space, since however no term
$\partial/\partial p$ appears, this motion is not a translation, it is rather like a $p$-dependent
 dilatation.\\
The factor $1/\omega_p^2$ in the commutation relation (\ref{omegaalg}) represents the curvature for the
sphere defined by $p_lp_l = p_0^2= {\mbox{fixed}}\not=0$. This sphere is 
however not a submanifold of the double cone $p_0=\pm \omega$.\\

\subsubsection{The differential operator X(x-conformal)}
Again, we assume $f$ to be a square integrable function $f$ from $\mathbb{R}^4 \rightarrow \mathbb{C}$.
The largest group of coordinate transformations in four-dimensional Minkowski spacetime which leaves
invariant the light cone $x^2=0$ is built up by the {\sl conformal} transformations: translations,
Lorentz
transformations, dilatations and special conformal transformations. Infinitesimally they operate
on $f(x)$ as follows
\begin{align}\label{358}
\delta^P_\mu &=\frac{\partial}{\partial x^\mu} \qquad \delta^M_{\mu\nu}
	       =x_\mu\frac{\partial}{\partial x^\nu}-x_\nu\frac{\partial}{\partial x^\mu}\\
\delta^D &=x^\lambda\frac{\partial}{\partial x^\lambda} \qquad 
		        \delta^K_\mu =(2x_\mu x^\nu-\delta_\mu^\nu x^2)\frac{\partial}{\partial x^\nu}
\end{align}							   
The Klein-Gordon operator $\Box$ turns out to be covariant with respect to these transformations,
hence on the formal level every solution transforms into a solution.\\ 
Defining operators 
\begin{align}\label{367}
	P_\mu &=-i\delta^P_\mu& \qquad M_{\mu\nu}=-i\delta^M_{\mu\nu}\\
	D     &=i(\delta^D+d)& \qquad K_\mu=-i(\delta^K_\mu +2dx_\mu)
\end{align}							   
we find the well-known conformal algebra
\begin{align}\label{confalg}
	[P_\mu,P_\nu]&=0\\
 [M_{\mu\nu},M_{\rho\sigma}] &=
 -i(\eta_{\mu\rho}M_{\nu\sigma}-\eta_{\mu\sigma}M_{\nu\rho}+\eta_{\nu\sigma}M_{\mu\rho}-\eta_{\nu\rho}M_{\mu\sigma} )\\
[P_\mu,P_\nu]&=0 \qquad  [M_{\mu\nu},P_\rho]=i(\eta_{\nu\rho}P_\mu-\eta_{\mu\rho}P_\nu)\\
[K_\mu,K_\nu]&=0 \qquad  [M_{\mu\nu},K_\rho]=i(\eta_{\nu\rho}K_\mu-\eta_{\mu\rho}K_\nu )\\
[D,M_{\mu\nu}]&=0\\
     [D,P_\mu]&=-iP_\mu \qquad [D,K_\mu]=iK_\mu\\
	[P_\mu,K_\nu]&=2i(\eta_{\mu\nu}D-M_{\mu\nu})
\end{align}
The commutation relations of $P$ with $K$ we interpret as preconjugation between
these operators; with
\be\label{385}
X(\mbox{x-conf})\equiv X(\mbox{conf}) \doteq K.
\ee
For later use we calculate the effect of $K_\mu$ on the Fourier transform:
\begin{align}\label{xconfFT}
	f(x)&=\frac{1}{(2\pi)^4}\int d^4p\, e^{-ipx} \tilde{f}(p)\nonumber\\
	K_\mu f(x)&=i\frac{1}{(2\pi)^4}\int d^4p\, e^{-ipx}
                	(-2(d-4)\frac{\partial}{\partial p^\mu}
				+2p^\lambda\frac{\partial^2}{\partial p^\lambda\partial p^\mu}
                       -p_\mu\frac{\partial^2}{\partial p^\rho\partial p_\rho})\tilde{f}(p)
\end{align}

Remark: In \cite{SwiecaVoelkel} it has been shown that conformal charges exist on
Hilbert space as essentially self-adjoint operators which then have at least one
self adjoint extension. The authors started from the well known off-shell special
conformal transformation, incorporated then the Klein-Gordon equation and derived
thereby the on-shell law for them. We shall find that form below in subsubsection
2.3.5.
Nevertheless we present the off-shell version here in the ``on-shell'' subsection
because the transition to the mass shell is smooth as will be explicitly shown
in section 4. 

\subsubsection{The differential operator X(p-conformal)}
If we solve the KG with vanishing mass parameter and consider it in Fourier
space we find
\be\label{FTm0}
f(x)=\frac{1}{(2\pi)^4}\int d^4p\, e^{-ipx} \delta(p^2){\hat{f}}(p).
\ee
The cone $p^2=0$ is left invariant by an infinitesimal conformal transformation 
in $p$-space. Hence we may define an operator $\nabla_\mu({\mbox{p-conf}})$ by
\be\label{nablapconf}
\nabla_\mu({\mbox{p-conf}})\tilde{f}(p)=(2p_\mu p^\lambda -\delta_\mu^
                \lambda p^2)\frac{\partial}{\partial p^\lambda}\tilde{f}(p),
\ee
and study its properties.\\
With the help of $\partial_\lambda\delta(p^2)=2p_\lambda\delta'(p^2)$ (the prime
indicates differentiation with respect to the argument) we find
\be\label{423}
\nabla_\mu({\mbox{p-conf}})\delta(p^2)=-2p_\mu\delta'(p^2),
\ee
i.e.\ this differential operator maintains $\delta(p^2)$ and thus  we may define an
operator $X_\mu({\mbox{p-conf}})$ and consider its action in $x$-space:
\begin{align}\label{430}
X_\mu({\mbox{p-conf}})\tilde{f}(p)&=i\nabla_\mu({\mbox{p-conf}})\tilde{f}(p)
	=i(2p_\mu p^\lambda-\delta_\mu^\lambda p^2)\frac{\partial}{\partial p^\lambda}\tilde{f}(p)\\
				   &=\int d^4p\, e^{ipx}
	(8\frac{\partial}{\partial x^\mu}+2x^\lambda\frac{\partial^2}{\partial x^\lambda\partial x^\mu}
	-x_\mu\Box )f(x).
\end{align}
By adding $2idp_\mu$ to $X_\mu({\mbox{p-conf}})$ we obtain
\begin{align}\label{pconfdiff}
	(X_\mu({\mbox{p-conf}})+2idp_\mu)\tilde{f}(p)&=i(\nabla_\mu({\mbox{p-conf}})+2p_\mu)\tilde{f}(p)\\
				   &=\int d^4p\, e^{ipx}
                                	(-2(d-4)\frac{\partial}{\partial x^\mu}
	                             +2x^\lambda\frac{\partial^2}{\partial^\lambda\partial^\mu}
	                             -x_\mu\Box )f(x).
\end{align}
This has precisely the form of $K_\mu f(x)$ in (\ref{xconfFT})!\\
Remark: The differential operator defined in (\ref{pconfdiff}) is an off-shell
operator.
It is treated here (in the section ``on-shell'') because it has an immediate
on-shell restriction and fits in here due to its geometric meaning.


\subsection{Preconjugate operators on one-particle Fock space}
The aim of the present subsection is to go over from one-particle wave functions to 
one-particle Fock space, thereby obtaining a basis independent formulation.
Most conveniently we start from the decomposition of a free scalar field into annihilation and creation
operators, which corresponds to separation into positive and negative frequency parts
\begin{align}\label{scalf+-}
\phi(x) &= \fp \int\! d^4p\, e^{-ipx}\tilde{\phi}(p)
	           = \fp \int\! d^4p\, e^{-ipx}\delta(p^2-m^2)\hat{\phi}(p)\\
	&= \fp \int\! \frac{d^3p}{2\omega_p}(e^{-ipx}a({\bf p})+e^{ipx}a^\dagger({\bf p}))
	           = \phi^{(+)}(x) + \phi^{(-)}(x).
\end{align}
Next we
have to translate the above differential operators acting on functions to operators
$X(a,a^\dagger)$ acting on one-particle Fock space states via the commutation relations
\footnote{Here we normalize the creation and annihilation operators Lorentz
covariantly, in contrast to \cite{Sibold,SiboldSolard,EdenSibold}}	
\be\label{cran-com}
[a({\bf p}),a^\dagger({\bf q})]=2\omega_p \delta^{(3)}({\bf p-q}) \qquad [a,a]=0=[a^\dagger,a^\dagger]
\ee
Eventually we aim at $X$'s which are charge like (i.e.\ map one-particle states into one-particle
states), Hermitian and -- most relevant for their final form -- reproduce the algebras we found for
the differential operators.

\subsubsection{The operator $X^{(\nabla)}(a^\dagger,a)$ on Fock space}
\noindent
We start from an operator $X^{(\nabla)}(\hbox{\rm pre})$ (``pre'' for ``preliminary'')
\begin{align}\label{Xnablapre}
X^{(\nabla)}_0(\hbox{\rm pre})=& i\alpha\int\!\frac{d^3p}{2\omega_p}\,
	(-\frac{\omega_p}{m^2})p^l\partial_l a^\dagger({\bf p})a({\bf p})\\
X^{(\nabla)}_j(\hbox{\rm pre})=& i\alpha\int\!\frac{d^3p}{2\omega_p}\,
	     (\frac{\partial}{\partial p^j} -\frac{p_j}{m^2}p^l\partial_l) a^\dagger({\bf p})a({\bf p})
\end{align}     
which implements the transformation law given by the differential operator $\nabla$ on
$a^\dagger$.\\
We check the Hermiticity of $X^{(\nabla)}(\hbox{\rm pre})$ via 
\be\label{XnablapreHerm}
\frac{1}{2}(X_\nu^{(\nabla)}(\hbox{\rm pre})\pm X_\nu^{(\nabla)\dagger}(\hbox{\rm pre}))= 
	\frac{i}{2}\int\!\frac{d^3p}{2\omega_p}(\alpha\nabla_\nu a^\dagger a\mp\bar{\alpha}a^\dagger\nabla_\nu a)
\ee	
(recall: $p_0=\omega_p, \partial/\partial p_0 \equiv 0)$)\\
and find for the transformation of $a^\dagger$
\be
[\frac{1}{2}(X_\nu^{(\nabla)}(\hbox{\rm pre})\pm X_\nu^{(\nabla)\dagger}(\hbox{\rm pre})),a^\dagger({\bf p})]=
-i(\frac{\alpha\pm\bar{\alpha}}{2}\nabla_\nu a^\dagger({\bf p})
\pm\frac{3\bar{\alpha}}{2}\frac{p_\nu}{m^2}a^\dagger({\bf p}))
\ee
For $\alpha\in \mathbb{R}$ the Hermitian part qualifies as a suitable extension to Fock space, whereas
the anti-Hermitian part does not play a role at this stage. (For $\alpha$ purely imaginary ``Hermitian''
and ``anti-Hermitian'' are interchanged; for $\alpha \in \mathbb{C}$ no suitable candidate is singled out.)\\

We therefore define now 
\be\label{Xnabla}
X^{(\nabla)}_\nu(a,a^\dagger)=\frac{i}{2}\int\!\frac{d^3p}{2\omega_p}\, (a^\dagger({\bf p})\nabla_\nu a({\bf p})
                                 -\nabla_\nu a^\dagger({\bf p})a({\bf p})).
\ee                                
This operator is charge like, (formally) Hermitian, made up from $a,a^\dagger$ and $\nabla$,
s.\ (\ref{Xnabladiff}), which
will guarantee that the mass shell constraint is maintained. Its most important property is however that
it reproduces the algebraic relation (\ref{PXnabladiff}) in the form
\begin{align}\label{PXnablacom}
[P_\mu, X^{(\nabla)}_\nu]&= i\int\!\frac{d^3p}{2\omega_p}(\eta_{\mu\nu}-\frac{p_\mu p_\nu}{m^2})a^\dagger({\bf p})a({\bf p})\\
                    &= i\eta_{\mu\nu}N-i\int\!\frac{d^3p}{2\omega_p}\frac{p_\mu p_\nu}{m^2}a^\dagger({\bf p})a({\bf p}),
\end{align}
 where $P_\mu$, $N$ denote the energy-momentum, resp.\ the number operator
 \be\label{PNdef}
P_\mu = \int\!\frac{d^3p}{2\omega_p}\,p_\mu a^\dagger({\bf p})a({\bf p}),\qquad
	 N = \int\!\frac{d^3p}{2\omega_p}\,a^\dagger({\bf p})a({\bf p}).
\ee
It therefore qualifies as an operator preconjugate to $P$ on Fock space.\\
To derive (\ref{PXnablacom}) becomes an exercise once one has calculated the ``transformation law'' for $a^{\#}({\bf p})$ 
(which stands for $a$ and $a^\dagger$)
\be\label{Xnabla-atransf}
i[X_\nu^{(\nabla)},a^{\#}({\bf p})]= \nabla_\nu a^{\#}({\bf p}) - \frac{3}{2}\frac{p_\nu}{m^2} a^{\#}({\bf p}).
\ee
It is to be noted that the second term does not contribute to the commutator (\ref{PXnablacom}),
since $P$ commutes with itself. It will also not push the field $\phi(x)$ off its mass shell --
just because it generates a translation in $x$-space. 
The remaining commutators which form the equivalent to (\ref{comXnabladiff}) on Fock space
read
\be \label{XXnablacom}
[X_\mu^{(\nabla)},X_\nu^{(\nabla)}] = \frac{i}{m^2}M_{\mu\nu}(a,a^\dagger),
\ee
with $M_{\mu\nu}$, the generators for Lorentz transformations in terms of creation and annihilation
operators (s.\ Appendix (A.2, A.3)).
With the help of (\ref{Xnabla-atransf}) one can also calculate the effect of $X^{(\nabla)}$ on the
field $\phi(x)$. The result reads
\be\label{XnablaPhi+-}
[X_\nu^{(\nabla)},\phi^{(\pm)}(x)]=\pm(x_\nu+\frac{1}{m^2}(\frac{3}{2}+x^\lambda\frac{\partial}{\partial x^\lambda})
                                     \frac{\partial}{\partial x^\nu})\phi^{(\pm)}(x),
\ee
and is interesting because it shows that Hermiticity of $X^{(\nabla)}$ leads to different behaviour of
positive and negative frequency part of the field.\\

\subsubsection{The operators $X(\mbox{wedge})$ on Fock space}
On the level of differential operators $X({\mbox{wedge}})$ can be obtained from
$X(\nabla)$ by a change of variables. On Fock space, however, one has to
introduce new creation and annihilation operators since both parts of
the mass hyperboloid contribute in a different way than in the standard
decomposition into positive and negative frequency parts. We postpone this
treatment to later work.\\
We also postpone the formulation of $X^{(<_0)}(a^\dagger,a)$ to later
work.\\

\subsubsection{The operator $X(\omega)$ on Fock space}
We proceed exactly as for $X^{(\nabla)}$. The differential operator yields first
\be\label{Xomegapre}
X^{(\omega)}_\nu(\hbox{\rm pre})= i\alpha\int\!\frac{d^3p}{2\omega_p}\,
	(-\frac{p_\nu}{\omega_p^2})p^l\partial_l a^\dagger({\bf p})a({\bf p})
\ee     
which implements the transformation law given by the differential operator $X^{(\omega)}$ on
$a^\dagger$.\\
Hermiticity:  via 
\be\label{XomegapreHerm}
\frac{1}{2}(X_\nu^{(\omega)}(\hbox{\rm pre})\pm X_\nu^{(\omega)\dagger}(\hbox{\rm pre}))= 
\frac{i}{2}\int\!\frac{d^3p}{2\omega_p}(\alpha\frac{p_\nu}{\omega_p^2} a^\dagger a
        \mp\bar{\alpha}a^\dagger\frac{p_\nu}{\omega_p^2} a)
\ee	
transformation of $a^\dagger$
\be
[\frac{1}{2}(X_\nu^{(\omega)}(\hbox{\rm pre})\pm X_\nu^{(\omega)\dagger}(\hbox{\rm pre})),a^\dagger({\bf p})]=
-i(\frac{\alpha\pm\bar{\alpha}}{2}\frac{p_\nu}{\omega_p^2} a^\dagger({\bf p})
\pm\frac{5i\bar{\alpha}}{2}\frac{p_\nu}{\omega_p^2}a^\dagger({\bf p}))
\ee
The reasoning is the same as for $X^{(\nabla)}$, we define therefore accordingly
\begin{align}\label{XomegaHerm}
	X^{(\omega)}_0=&\frac{i}{2}\int{ }\frac{d^3p}{2\omega_p}\frac{1}{\omega_p}
	\left(-a^\dagger p^l\partial_l a + p^l\partial_l a^\dagger a)\right)\\
	X^{(\omega)}_j=&\frac{i}{2}\int{ }\frac{d^3p}{2\omega_p}\frac{p_j}{\omega_p^2}
	\left(a^\dagger p^l\partial_l a - p^l\partial_l a^\dagger a)\right)
\end{align}
and remark that $X^{(\omega)}_\mu$ is the Lorentz vector whose zeroth component coincides with
$X^{SiSo}_0$ 
It entails $a^\dagger$ with the transformation law
\be
[X^{(\omega)}_\mu,a^\dagger(\mathbf{p})]=
               -i(\frac{p_\mu}{\omega_p^2}p^l\partial_l a^\dagger(\mathbf{p})
                         +\frac{5}{2}\frac{p_\nu}{\omega_p^2}a(\mathbf{p})^\dagger).
\ee
For the algebra one finds
\begin{align}
[P_\mu,X^{(\omega)}_\nu] =&i\int{ }\frac{d^3p}{2\omega_p}\frac{p_\mu p_\nu}{\omega^2_p}
                                        	a^\dagger a\\
[X^{(\omega)}_\mu,X^{(\omega)}_\nu]=&0
\end{align}						
We recall that only the zeroth component of $X(\omega)$ transforms as (on-shell) Lorentz vector.\\
The commutator $[P_0,X_0]=iN$ was in \cite{SiboldSolard} the reason for combining $X_0$
with $X_j$'s having the commutator $-iN$ with $P_j$. Those did however not have
$X^{(\omega)}_0$ as its zeroth component, but transformed into a generic tensor under Lorentz.
Here we find that $X_j$'s do transform into a generic tensor once we extend $X_0$ to $X_j$'s
by demanding Lorentz covariance for it.\\

\subsubsection{The operator X(x-conformal) on Fock space}
In \cite{SiboldSolard,EdenSibold} the Fock space representatives for these transformations have been derived
by starting from the conformal current as $x$-moment of the improved energy-momentum tensor
and then going over to the respective charge, first in $x$- and thereafter in $p$-space.
Equally well one can derive them by translating the off-shell $x$-space variation of a scalar
field into $p$-space, go on-shell
and follow then the derivation
pattern used for $X^{(\nabla)}$ and $X^{(\omega)}$. The result is in both cases the same
and reads (in covariant normalization of the
annihilation and creation operators):
\begin{align}\label{confcharge0}
	K_0=&\int{}\frac{d^3p}{2\omega_p}~\omega_p~a^\dagger(\mathbf{p}) \partial^l\partial_l a(\mathbf{p}) \\ \label{confchargej}
	K_j=&\int{}\frac{d^3p}{2\omega_p}~a^\dagger(\mathbf{p})\left(p_j\partial^l\partial_l
-2p^l\partial_l\partial_j-2\partial_j\right)a(\mathbf{p}) 
\end{align}
Remarkably they come out as Hermitian operators immediately.
They yield as transformation law
\begin{align}\label{xconftransa1}
	[K_0,a^\dagger(\mathbf{p})]=& \omega_p\partial^l\partial_l a^\dagger(\mathbf{p})\\
	\label{xconftransa2}
	[K_j,a^\dagger(\mathbf{p})]=&\left(p_j\partial^l\partial_l
	                  -2p^l\partial_l\partial_j-2\partial_j\right)a^\dagger(\mathbf{p})
\end{align}
for the creation operator.\\
$K_\mu$ forms together with $P_\mu, M_{\mu\nu}, D$ the conformal algebra
(\ref{confalg}) realized on Fock space.

\subsubsection{The operator X(p-conformal) on Fock space}
In the, by now standard, procedure we start from the differential operator (\ref{nablapconf})
and (\ref{pconfdiff}) (by some abuse of notation)
\be
\nabla_\mu({\mbox{p-conf}})\tilde{f}(p)=((2p_\mu p^\lambda -\delta_\mu^
                             \lambda p^2)\frac{\partial}{\partial p^\lambda}+2dp_\mu) \tilde{f}(p),
\ee
i.e.\ off-shell ($d \in \mathbb{R})$. On-shell ($p^2=0, p_0=\omega_p, \partial/\partial p^0\equiv0$) we define
\be \label{Xpconfpre}
K^{({\rm p})}_\nu({\mbox{pre}})= \alpha\int{ }\frac{d^3p}{2\omega_p}(2p_\nu(d+
                                              p^l\frac{\partial}{\partial p^l}))a^\dagger a, 
\ee
and calculate
\be \label{XpconfpreHerma}
\frac{1}{2}(K^{({\rm p})}_\nu({\mbox{pre}})\pm K^{({\rm p})\dagger}_\nu({\mbox{pre}}))
= \frac{1}{2}\int{ }\frac{d^3p}{2\omega_p}(2(d(\alpha+\bar{\alpha})\mp 3\bar{\alpha})p_\nu a^\dagger a
+(\alpha\mp\bar{\alpha})p_\nu p^l\frac{\partial}{\partial p^l}))a^\dagger a. 
\ee
Hence we find for $\alpha=\bar{\alpha}\in \mathbb{R}$
\begin{align}\label{XpconfHer}
	K^{({\rm p-Herm})}_\nu\equiv &\frac{1}{2}(K^{({\rm p})}_\nu({\mbox{pre}})
       +K^{({\rm p})\dagger}_\nu({\mbox{pre}}))
=2(d-\frac{3}{2})\alpha P_\nu\\
K^{({\rm p-anti-Herm})}_\nu\equiv &\frac{1}{2}(K^{({\rm p})}_\nu({\mbox{pre}})-
K^{({\mbox{p}})\dagger}_\nu({\mbox{pre}})) =\alpha(\frac{3}{2}-i\check{D})P_\nu,\\
	{\mbox{with}}\quad \check{D}\equiv &ip^l\frac{\partial}{\partial p^l}.
\end{align}
Again, for $\alpha$ purely imaginary the roles of ``Hermitian'' and ``anti-Hermitian''
are interchanged, for $\alpha \in \mathbb{C}$ the outcome is a general mixture.
We conclude that the case with real $\alpha$ is the only reasonable one and that
(for $d\not=3/2$) the operator $K^{(P-Herm)}$ is equivalent to the standard translations
$P$. Hence we need not pursue it further.\\

\newpage
\section {Off-shell approach: Green functions}
\noindent
In the present section we shall work with {\sl Green} functions as the relevant system of 
functions on which the preconjugate operators act. The Green functions are defined as the vacuum
expectation value of time ordered products of fields
\begin{equation}
	G_n(x_1,...,x_n) = <0|T(\phi(x_1)...\phi(x_n))|0>.
\end{equation}	
(For simplicity of notation we have chosen a scalar field as example.) The field is an operator, hence
the c-number $G_n$ is, more mathematically speaking, a distribution -- a fact which has to be taken
into account in the sequel.\\
The Green functions can be obtained from a generating functional $Z(J)$
\begin{eqnarray}
	G_n(x_1,...x_n) &=& \frac{\delta}{i\delta J(x_1)}\cdots\frac{\delta}{i\delta J(x_n)} Z(J)_{|J=0}\\
	  Z(J)          &=& T<0|exp(i \int\,d^4x J(x) \phi(x))|0>
\end{eqnarray}
(Unless one supplies $Z(J)$ with true content this definition is purely formal.)\\
On suitable test function spaces the Green functions can be Fourier transformed
\begin{equation}
	G_n(x_1,...,x_n) = \frac{1}{(2\pi)^n}\int\,d^{4n}p e^{-i(p_1x_1+\cdots+p_nx_n)}\tilde{G}(p_1,...,p_n) ,
\end{equation}	
such that the considerations on functions above apply.\\
In the context of the Lehmann-Symanzik-Zimmermann (LSZ) formulation of
QFT one is able to derive Hilbert space operators, in fact Fock space
operators, from off-shell Green functions via the so-called reduction
formalism. The preconjugate operators which we wish to construct
can thus be formulated as functional differential operators acting on $Z$.
A familiar case is given by those which formulate tentative symmetries
like conformal. Hence we study first how preconjugate pairs can be
realized by Ward identity like functional differential operators and 
second, which ones can be continued on-shell.\\

\subsection{A first trial}
Let $\Gamma$ be the generating functional for vertex functions
\begin{equation}
	\Gamma = \sum \frac{1}{n!}\int\,d^{4n}x\,\varphi(x_1),...,\varphi(x_n)\Gamma_n(x_1,...,x_n).
\end{equation}
\noindent
We introduce functional differential operators which represent
field transformations. For translations and $X$-transformations resp.\ they read
\begin{eqnarray}
W^T_\mu \Gamma &=& i\int \!d^4 x\; \partial_\mu \varphi(x) \frac{\delta}{\delta \varphi(x)}\Gamma\\
W^X_\nu \Gamma &=& \int \!d^4 y\; y_\nu \varphi(y) \frac{\delta}{\delta \varphi(y)}\Gamma.
\end{eqnarray}
Their commutator is easily calculated:
\begin{eqnarray}
\left[W^T_\mu,W^X_\nu\right] &=& -i\eta_{\mu\nu}\; \mathcal{N}\\
\hbox{\rm with}\qquad \mathcal{N}&=& \int \!d^4 x\; \varphi(x)\frac{\delta}{\delta\varphi(x)}
\end{eqnarray}
When applied to an n-point vertex function, $\mathcal{N}$ yields $n$, it counts the number
of legs of a respective Feynman diagram.  This is fine
off-shell, however not satisfactory on-shell, as we show now.\\
By Legendre transforming $\Gamma$ we introduce first $Z_c(j)$ -- the generating functional for connected
Green functions, second by exponentiation the generating functional for general Green functions:
\begin{eqnarray}
Z_c(j) &=& \Gamma(\varphi(j)) - \int\! dx\; j(x)\varphi(j(x))
                                             \qquad\frac{\delta\Gamma}{\delta\varphi(x)} = -j(\varphi(x))\\
Z(j) &=& e^{iZ_c(j)}
\end{eqnarray}
Standard perturbation theory proceeds recursively with the number of closed loops in Feynman diagrams
as expansion parameter and in the tree approximation (number of loops equal to zero) $\Gamma$ can be identified
with the classical action. For its proper definition one has to refer to a renormalization
scheme which we chose her to be the momentum space subtraction scheme of Bogoliubov, Para\v{s}iuk,
Hepp, Zimmermann (BPHZ). Within this scheme the effect of differential operators on the functionals
$\Gamma, Z_c, Z$ can be conveniently expressed with the help of the action principle:
\begin{eqnarray}
\varphi(x)\frac{\delta}{\delta\varphi(x)}\Gamma &=&
                     \left[\varphi(x)\frac{\delta}{\delta\varphi(x)}\Gamma_{eff}\right]\cdot\Gamma\\
-j(x)\frac{\delta}{\delta j(x)}Z_c &=& \left[\varphi(x)\frac{\delta}{\delta\varphi(x)}\Gamma_{eff}\right]\cdot Z_c\\
ij(x)\frac{\delta}{\delta j(x)}Z &=& \left[\varphi(x)\frac{\delta}{\delta\varphi(x)}\Gamma_{eff}\right]\cdot Z\\
\end{eqnarray}
Here $\Gamma_{eff}$ is given by the sum of the classical action and all counter terms.\\
This version of the action principle holds for all linear field transformations (after assigning suitable
subtraction degrees to the normal products $\left[...\right]$).\\
Let us now use the action principle in the context of a self-interacting, massive scalar field with
\begin{equation}
\Gamma_{eff} = \int \! d^4 x\;(\frac{1}{2}(\partial\varphi\partial\varphi - m^2\varphi^2) 
                                                            - \frac{\lambda}{4!}\varphi^4) + \Gamma_{counter}. \\
\end{equation}							    
Suppressing for simplicity of writing the contribution from the counterterms we find 
\begin{equation}
w^{\mathcal{N}}(x)\Gamma\equiv \varphi(x)\frac{\delta}{\delta\varphi(x)}\Gamma =
              \left[\varphi(x)\left(-(\Box+m^2)\varphi(x)-\frac{\lambda}{3!}\varphi(x)^3\right)\right]\cdot\Gamma.
\end{equation}
Here we have introduced $w^{\mathcal{N}}(x)$ which yields upon integration $\mathcal{N}$.\\
On Z the respective equation reads
\begin{equation}
w^{\mathcal{N}}(x)Z\equiv ij(x)\frac{\delta}{\delta j(x})Z =
              \left[\varphi(x)\left(-(\Box+m^2)\varphi(x)-\frac{\lambda}{3!}\varphi(x)^3\right)\right]\cdot Z.
\end{equation}	      
The left hand side of this equation represents contact terms which vanish once we apply LSZ reduction.
It yields the operator equation
\begin{equation}
\left[\varphi(x)\left(-(\Box+m^2)\varphi(x)-\frac{\lambda}{3!}\varphi(x)^3\right)\right]^{Op} = 0.
\end{equation}
It is quite meaningful: it constitutes the socalled bilinear field equation, but it implies
after integration that on the operator level $\mathcal{N}$ vanishes. Hence $\mathcal{N}$ \. is certainly not
what we look for in our search for a coordinate operator.\\
In order to have some guidance as to how we should proceed we have a glance at the construction of the
energy-momentum operator $P_\mu$. $W_\mu^T$ reads on Z as follows:
\begin{equation}
W_\mu^T Z \equiv -i\int\! d^4 x\; \partial_\mu j(x)\frac{\delta}{\delta j(x)}Z = 0.
\end{equation}
It expresses translation invariance of our system. On the non-integrated level we find 
\begin{equation}
w_\mu^T(x) Z \equiv -i\partial_\mu j(x)\frac{\delta}{\delta j(x)}Z =
                                                 \left[-\partial^\nu T_{\mu\nu}(x)\right]\cdot Z.
\end{equation}
LSZ reducing this latter result we obtain the operator equation
\begin{equation}
\partial^\nu T^{op}_{\mu\nu}(x)=0,
\end{equation}
the energy-momentum tensor (EMT) is conserved as an operator.\\
Introducing the energy-momentum vector by
\begin{equation}
P_\mu = \int \! d^3 x \; T_{\mu 0}(x),
\end{equation}
the conservation equation of the EMT tells us that, by LSZ reduction, $P_\mu^{op}$
is a time independent operator
and furthermore, that it generates the translations on the operator field
$\varphi^{op}(x)$:
\begin{equation}
\left[P_\mu^{op},\varphi^{op}(x)\right] = -i\partial_\mu\varphi^{op}(x). 
\end{equation}
Comparing with the situation for $\mathcal{N}$, the difference clearly originates from the fact that the
conservation equation for the EMT permits the definition of $P_\mu$ as a non-trivial operator
on-shell ensuing a non-trivial transformation for the field operator $\varphi^{op}(x)$.
Hence we conclude that we should search for a replacement of $\mathcal{N}$ by a functional differential
operator whose unintegrated representative yields a total divergence. The same arguments apply to
the above $W^X_\nu$ because it also does not lead to field transformation on-shell, ie.\ for the field 
operator.\\

\subsection{Systematic search}
In our first trial to find $W^X$'s we were lead: first by a dimensional argument -- the integrand of $W^X$ must have
dimension -1, and second by the desire to find $\eta_{\mu\nu}$ on the r.h.s.\ of the conjugation commutator.
Eventually we have indeed to have dimension -1, but we certainly can admit a more complicated r.h.s.
So, in a fairly general ansatz we assume
\begin{equation}
\tilde{w}^T_\mu \Gamma = \partial_\mu\varphi(x)\frac{\delta}{\delta \varphi(x)}\Gamma 
                          - \frac{1}{4}\partial_\mu\left(\varphi(x)\frac{\delta}{\delta \varphi(x)}\Gamma\right),
\end{equation}			  
i.e. add a very specific total derivative to the first term, which integrated generates the global
translations as before, whereas the second term contributes locally to the potential conservation
equations. Clearly the integral of a second $x$-moment of the second term produces a contribution
of type $W^X_\mu$ which we had before, so we are on the right track. The most remarkable property
of the contact terms $\tilde{w}^T_\mu$ is however that they close under commutation
\cite{KrausSibold1} (eq. (2.15)):
\begin{equation}\label{locclosure}
\left[\tilde{w}^T_\mu (x),\tilde{w}^T_\nu (y)\right] =
          \partial_\nu^y\delta(x-y)\tilde{w}^T_\mu(y)-\partial_\mu^x\delta(x-y)\tilde{w}^T_\nu(x).
\end{equation}
Hence $x$-moments of $\tilde{w}^T$ will have well-defined transformation properties.
Since, in fact, (\ref{locclosure}) is the algebra for general coordinate 
transformations and its implementation on the level of fields generates the 
diffeomorphisms on fields \cite{KrausSibold3}, i.e.\ general relativity, we are
sure, that we are dealing with the most general possibility of enlarging $W^X$'s. 
However, for extending such properties to operators we have to have recourse to
a specific $\Gamma$, i.e.\ in the tree approximation to specific actions.
Acting with $\tilde{w}^T_\m$ on an action invariant under translations we certainly obtain 
$\partial^\nu T_{\mu\nu}$ of some EMT $T_{\mu\nu}$, yielding $P_\mu$ as before. Contracting
$\tilde{w}^T_\lambda$ with some moment $a_\mu^\lambda(x)$ and integrating we obtain first of all
moment contact terms but secondly by operating on the same action as before we get moments of 
the EMT . Let us be specific.\\
We work in the tree approximation, hence with the action of eq.\ (11), at $m=0$ (the reason for this
restriction will become clear below).
The local conservation eq.\ for the translations reads
\begin{eqnarray}
\tilde{w}^T_\mu(x)\Gamma^{(0)}&=& -\partial^\lambda T_{\mu\lambda}\\
\hbox{\rm with} \qquad T_{\mu\lambda} &\equiv& T_{\mu\lambda}^{\rm impr}\\
    &=& \partial_{\mu}\phi\partial_{\lambda}\varphi - \frac{1}{2}\eta_{\mu\lambda} \partial\varphi \partial\varphi
	 - \frac{1}{4}\eta_{\mu\lambda} \varphi \Box\varphi - \frac{1}{6}(\partial_{\mu}\partial_{\lambda}-
	                                      \eta_{\mu\lambda}\Box)\varphi^2
\end{eqnarray}
Choosing as moments $a_\mu^\lambda(x) = 2x_\mu x^\lambda-\eta_\mu^\lambda x^2$
one finds
\begin{eqnarray}\label{moments}
w^K_\mu\Gamma^{(0)}\equiv a_\mu^\lambda(x)\tilde{w}^T\lambda(x)\Gamma^{(0)}=
                          \partial^\rho K_{\mu\rho}(x)\\
\hbox{\rm with} \qquad K_{\mu\rho} = a_\mu^\lambda T_{\rho\lambda}(x)
\end{eqnarray}
Hence, these equations give us conserved currents: for the translations $T_{\mu\nu}^{\rm impr}$
and for the special conformal transformations $K_{\mu\rho}$ (here $m=0$ becomes relevant). The
integrated contact terms
are indeed those of translations and special conformal transformations of a scalar field
with canonical dimension 1:
\begin{eqnarray}
W^T_\mu&\equiv& i\int \! d^4 x \;\partial_\mu \varphi (x) \frac{\delta}{\delta\varphi(x)}\\
W^K_\nu&\equiv& i\int \! d^4 x \;((2x_\nu x^\lambda -\eta_\nu^\lambda x^2)\partial_\lambda
                      \varphi\frac{\delta}{\delta \varphi}+2x_\nu \varphi\frac{\delta}{\delta\varphi})
\end{eqnarray}
The algebra of these integrated WI operators can either be calculated directly or obtained 
from the closure of the local differential operators (\ref{locclosure}) by multiplying with the moment
function and integrating. It is the well-known conformal algebra
\begin{eqnarray}
  \left[W^T_\mu,W^K_\nu\right] &=& 2i(\eta_{\mu\nu}W^D - W^M_{\mu\nu})\\
    \hbox{\rm with} \qquad W^D &\equiv& i\int \! d^4 x \;(1+x\partial_x)
	                          \varphi \frac{\delta}{\delta\varphi}\\
\hbox{\rm and } \qquad W^M_{\mu\nu} &\equiv& i\int \! d^4 x \;(x_\mu\partial_\nu 
	             - x_\nu\partial_\mu)\varphi \frac{\delta}{\delta\varphi}
\end{eqnarray}							       
$W^D$: dilatations; $W^M_{\mu\nu}$: Lorentztransformations.\\
The first lesson we learn from this example is the following. The choice $X_\mu = K_\mu$ is certainly
good. The currents are local operators which are conserved. $K_\mu$ as the charge of a conserved current,
generates the transformation and is conserved in time. It is a Lorentz vector. The transition from
off-shell to on-shell is
possible. However its algebra is not of the standard canonical (Hamiltonian) form. This problem
will be discussed in \cite{PottelSiboldII}.\\
For the second lesson we look separately at every term coming from the moment function.
Every one of them has the right dimension, but none of them yields a conserved current (separately).
This is well-known \cite{CallanColemanJackiw}: admitting in (\ref{moments}) for $a_\mu$ an
arbitrary function and asking for a conserved current one finds that only the first
and second $x$-moments of the EMT yield conserved currents. They lead to dilatations,
Lorentz transformations and special conformal transformations respectively.\\
In reference \cite{EdenSibold} the generators $K_\mu$ have been expressed for the free massless scalar field
in terms of particle creation and annihilation operators. Remarkably, it turned out that they are
of pure charge type: they contain only bilinear products of type $a^\dagger a$, but not of type
$aa$ or $a^\dagger a^\dagger$. It was found that the origin of this property was the improvement
of the EMT (a consequence of the second term in (\ref{locclosure})). Since a mass term would break the conformal symmetry, only the massless model will
deliver a good candidate for the preconjugate operator $X_\mu$.\\
Let us summarize this section. The only preconjugate operator $X_\mu$ which:\\
 - is local in $x$-space,\\
 - conserved in time,\\
 - permits a transition off-shell/on-shell,\\
 - is charge-like in Fockspace,\\
 - transforms as a Lorentz vector\\ 
 is $X_\mu = K_\mu$, the charge generating special conformal
 transformation.\\
 This holds true for the free theory and also for the tree approximation
 of the interacting theory. In higher orders the conformal anomaly will cause 
 effects to be understood.\\
 This result implies that all other on-shell operators $X_\mu(a^\dagger,a)$
 which we  defined above in the  on-shell section, are non-local in field space
 and/or lack Lorentz covariance. We remind the reader
 that we are working on four dimensional flat Minkowski spacetime.\\

 \section{Extension, Applications}

         \subsection{(Anti-)deSitter space}
In the previous section the underlying spacetime was ordinary four-dimensional Minkowski
and a preconjugate partner for the translations was found in a generator of the conformal
group $SO(2,4)$. In detail, we chose a representation of $SO(2,4)$ which contains the 
Poincar\'e group $ISO(1,3)$ as a subgroup and get Minkowski space
$\mathbb{M}=ISO(1,3)/SO(1,3)$ by taking the quotient with the Lorentz group $SO(1,3)$.
Then translations are singled out to be the candidate for finding (pre-)conjugate partners.
Now we observe that we obtain four-dimensional (Anti-)deSitter space
$dS_4 = SO(1,4)/SO(1,3)$ ($AdS_4=SO(2,3)/SO(1,3)$, resp.) again by choosing a subgroup of
the of the conformal group and taking the quotient with the Lorentz group. We will sketch
now, how the resulting four generators of the homogeneous spaces are related with each other
and thus previous considerations can be extended to (Anti-)deSitter space.\\
In the theory of Lie algebras, it is well known that certain classes of algebras can be
obtained from each other by deformation \cite{Gerstenhaber:1963zz} and contraction
\cite{Inonu:1953sp}. For simplicity, we focus on the latter where a scaling parameter
$\epsilon$ is introduced such that in the limit $\epsilon \rightarrow 0$ the simple 
Lie-algebra is transformed into a semisimple Lie-algebra. \\
More precisely, we start by specifiying the conformal algebra $so(2,4)$ with generators
$J_{ab}$ where $a,b\in\{-1,0,...,4\}$.
\begin{align}
 -i[J_{ab},J_{cd}] & = \eta_{ad} J_{bc} + \eta_{bc} J_{ad} - \eta_{ac} J_{bd} 
 	                      - \eta_{bd} J_{ac} \\
              \eta & = \mathrm{diag}(+1,+1,-1,-1,-1,-1)
\end{align}
where $\mu\in\{0,...,3\}$. \\
For the introduction of the parameter $\epsilon$, we follow the ideas of
\cite{Aldrovandi:2006vr} and assign a physical parameter to $\epsilon$, namely the inverse
of squared spacetime radius $r^{-2}$ or equivalently the scalar curvature $R$. Regarding 
the physical dimension of the generators, we have $P_\mu\mapsto 2rP_\mu$ and
$K_\nu\mapsto \frac{1}{2r}K_\nu$ and we use two arguments that enable us to perform the
limit in both spaces simultaneously. In order to obtain a well-defined limit, one has to 
perform a compactification of the spaces, which leads to an additional factor $1/r$ for 
each generator $J_{4\mu}$.
\begin{align}
I_{4\mu} = P_\mu - \frac{\epsilon}{4} K_\mu
\end{align}
Indeed, the contraction limits of deSitter and Anti-deSitter space are opposite, i.e.
while the first requires $r^2\rightarrow\infty$, the latter requires $r^2\rightarrow0$.
Using the automorphism $P_\mu\rightarrow K_\mu$ and $K_\mu\rightarrow -P_\mu$ in the
conformal algebra $so(2,4)$, we relate $J_{-1,\mu}$ to $J_{4,\mu}$ and arrive at
\begin{align}
I_{-1,\mu} = \frac{\epsilon}{4} P_\mu + K_\mu.
\end{align}
Now performing the limit, we obtain
\begin{align}
  \lim_{\epsilon\rightarrow0} I_{4\mu} & = P_\mu \\
\lim_{\epsilon\rightarrow0} I_{-1,\mu} & = K_\mu 
\end{align}
and in particular
\begin{align}
\lim_{\epsilon\rightarrow0} [I_{4\mu},I_{-1,\nu}] = 2i(\eta_{\mu\nu} D - M_{\mu\nu}).
\end{align}
Hence for every spacetime which has an isometry group in $SO(2,4)$, we may consider the
generators $J_{4,\mu}$ and $J_{-1,\nu}$ as candidates for (pre-)conjugate partners.

\subsection{Applications}
In \cite{Muc1} we investigated how to obtain a non-commutative spacetime from
coordinate operators. The investigation in this case was guided by physical
intuition. On the one hand we deformed the quantum mechanical coordinate
operators by their conjugate momenta employing some deformation matrix.
This resulted into Moyal-Weyl commutation relations for the deformed 
coordinates. On the other hand we deformed the free Hamiltonian with
the ordinary coordinates and a respective deformation matrix. By identifying
suitably the parameters we obtained a physical interpretation for them.
E.g.\ in the case of Landau quantization the deformation parameters
represent the external magnetic field applied and the deformed coordinates
constitute the guiding center coordinates which are measurable. Ordinary
coordinates together with external magnetic field can be understood as basing
a free Hamiltonian on a Moyal spacetime.\\
In the present context another result derived in \cite{Muc4} is more
relevant: even for a {\sl preconjugate} pair like $P\mu$ and $K_\nu$, 
energy-momentum and special conformal operators in QFT, one can 
apply the deformation technique in the same spirit. Deforming the coordinates
on which a
(scalar) quantum field depends with the special conformal operators
and a deformation matrix one obtains for these coordinates a non-commutative
spacetime. The originally local field becomes a {\sl wedge-local}
field.\\
In some detail one proceeds as follows.
Within the framework of warped convolutions \cite{BLS},
the deformed associative product  $\times_{\theta}$  of
$A,B$ is defined as
\begin{equation}
   A\times_{\theta}B=(2\pi)^{-d}\iint d^dv d^du e^{-ivu}
             \alpha_{\theta v}(A)\alpha_{u}(B).
\end{equation}
Here $\alpha$ denotes the adjoint action 
$\alpha_{\theta v}(A)=U(\theta v)AU^{-1}(\theta v)$ with 
$U(v)=\exp{(iv_\mu G^\mu)}$ and $G$ being the generator of the deformation.
Furthermore, the deformed commutator $[A\stackrel{\times_{\theta}}{,}B]$ of A,
B reads
\begin{equation}\label{dc}
[A\stackrel{\times_{\theta}}{,} B] :=A\times_{\theta}B-B\times_{\theta}A.
\end{equation}
In analogy one obtains for $A,B$ being coordinates $x$ and $G$ being special
conformal transformations
\begin{align*}
	[x_{\mu}\stackrel{\times_{\theta}}{,} x_{\nu}]&=
	-2i(\theta_{\mu\nu}x^2
	+2\left((\theta x)_{\mu}x_{\nu}-(\theta x)_{\nu}x_{\mu} 
	\right)x^{2}
\end{align*}
up to third order in the deformation parameter $\theta$.
This corresponds to a non-constant, non-commutative spacetime.\\
For the other preconjugate operators $X(\nabla), X(<), X(\omega)$ the
precisely analogous construction does not seem to be possible since they
act non-locally on a quantum field. For them other applications have to
be found.\\

\newpage

\section{Discussion and conclusions}
The present search for preconjugate pairs of operators has been guided first by
geometry:
identifying tangential derivatives of the mass hyperboloid $p^2=m^2$ we were lead to
differential
operators on one-particle wave functions -- we obtained $X(\nabla)$. Since in the
massless limit the hyperboloid degenerates into a double cone, which is only a
topological
manifold, tangential derivatives lead to $X(<_0)$ and $X(\omega)$, also realized as 
differential operators on functions. Since the double cone $p^2=0$ is invariant
under
conformal transformations we found $X(\hbox{\rm p-conformal})$. The invariance of
the massless Klein-Gordon equation under conformal transformations in $x$-space
which is related to the invariance of $x^2=0$ lead us to the differential operators
$X(\hbox{\rm x-conformal}) \equiv K$ again acting on functions. We conclude that
on this level we have found all differential operators which are characterized
by intrinsic properties of the underlying structures, i.e.\ the mass hyperboloid,
resp.\ the double cones (in $p$- and $x$-space).\\
In a second step these differential operators have been represented on Fock-space,
here realized
via charge-like operators composed of creation and annihilation operators of a real
scalar field. Whereas the respective
algebras are identical, the variations of creation/annihilation operators do not
in all cases agree with the application of the respective differential operators
on functions.\\

\noindent
In order to find out which on-shell operators $X$ can be derived as {\sl local}
operators from insertions into Green functions we studied the appropriate
candidates in sect.\ 3. The result that the ordinary space-conformal 
$X=K$ is singled out by this requirement is important, because in turn
this shows that all other $X$'s are non-local in space, hence would not
qualify as ``local observable'' in the traditional sense of the word.
They would not appear in the algebra of observables \`a la \cite{Haag}.\\

\noindent
The extension to a more general setting as far as spacetime is concerned
seems to be possible, we examplified this by having a look at (Anti-)deSitter
in subsection 4.1.
As an application of the preconjugate operator $K$ we refered to the
construction of an non-constant, non-commutative spacetime in subsection
4.2.\\

\noindent
Generalizations to non-zero spin and to gauge theories is in principle
straightforward. Since conformal invariance is well studied in the literature
it should be possible to formulate that case first and then in analogy to it
the example of, say $X(\nabla)$ and its transition to $Q(\nabla)$. Similarly
it seems quite feasible to treat supersymmetric models.\\

\noindent
One last point to be discussed concerns the effects of interaction. All of
our $X(a,a^\dagger)$'s operate in the asymptotic region. All, but $K$, 
are non-local hence would transform the $S$-matrix non-trivially and in a
controllable way, since they are charge-like. It is however not obvious
how to construct their interacting versions (of which they would have
to be the asymptotic limit). The only one for which this is possible
in principle is $X=K$. In an interacting $\Phi^4$-theory the conformal
charge is not conserved, however, due to the well-known conformal anomaly.
In this case one can control the deviation from symmetry by introducing
an external field given as a local coupling, (s. \cite{PottelSibold}).
It is far from obvious how one had to incorporate and interpret this
phenomenon.\\

Acknowledgements\\
A.M.\ would like to thank Andreas Andersson for many discussions.
S.P.\ gratefully acknowledges financial support by the Max Planck Institute
for Mathematics in the Sciences and its International Max Planck Research
School (IMPRS) ``Mathematics in the Sciences''. \\


\section{Appendix: Charges of the conformal group}
For convenience of the reader we present the generators of the conformal group in
terms of creation and annihilation operators (in covariant normalization) on
Fock space.\\
\[
	\begin{array}{lclr}
P_{\mu}&=&\int{}d\mu~p_{\mu}a(\mathbf{p})^{\dagger}a(\mathbf{p})&\hfill (A.1)\\
M_{j0}&=&-i\int{}d\mu~\omega_p a({\mathbf p})^{\dagger}\partial_j a(\mathbf{p})
		&\qquad \qquad (A.2)\\
M_{jk}&=&i\int{}d\mu \left(p_j\partial_k a(\mathbf{p})^\dagger
	a(\mathbf{p})-p_k\partial_j a(\mathbf{p})^\dagger a(\mathbf{p})\right)
		&\qquad\qquad (A.3)\\
	D&=&-i\int{}d\mu~\left(a(\mathbf{p})^\dagger a(\mathbf{p})
+a(\mathbf{p})^\dagger p^l\partial_l a(\mathbf{p})\right)& \hfill (A.4)\\
	K_0&=&\int{}d\mu~\omega_p~a(\mathbf{p})^\dagger \partial^l\partial_l
	a(\mathbf{p}&\qquad \qquad (A.5)\\
	K_j&=&\int{}d\mu~a(\mathbf{p})^\dagger
 \left(p_j\partial^l\partial_l-2p^l\partial_l\partial_j-2\partial_j\right)
 a(\mathbf{p}) &\qquad\qquad (A.6)\\
\end{array}
\]
Here the measure reads $d\mu=d^3\,p/{2\omega_p}$ and the relation to
the non-covariant normalization as used in \cite{SiboldSolard},
\cite{EdenSibold}
is given by $a_{\hbox{\rm non-cov}}= a/\sqrt{2\omega_p}$.

\bibliographystyle{phjcp}
\bibliography{bphzl}
\end{document}